%% file: main.tex
\documentclass[10pt,twocolumn,letterpaper]{article}

\usepackage{iccv}              


\definecolor{iccvblue}{rgb}{0.21,0.49,0.74}
\usepackage[pagebackref,breaklinks,colorlinks,allcolors=iccvblue]{hyperref}
\usepackage{pifont}
\usepackage{makecell, multirow, multicol}

\def\paperID{12470} 
\def\confName{ICCV}
\def\confYear{2025}

\title{Audio-visual Event Localization on Portrait Mode Short Videos}

\author{Wuyang Liu$^2$  Yi Chai$^2$  Yongpeng Yan$^2$  Yanzhen Ren$^{1,2,*}$\\
$^1$Key Laboratory of Aerospace Information Security and Trusted Computing, Ministry of Education\\
$^2$School of Cyber Science and Engineering, Wuhan University\\
{\tt\small \{liuwuyang, chaiyi, yanyongpeng, renyz\}@whu.edu.cn}
}

\begin{document}
\maketitle
\input{sec/0_abstract}    
\input{sec/1_intro}
\input{sec/2_related_works}
\input{sec/3_ave_pm}

\input{sec/4_experiments}
\input{sec/5_recipes}
\input{sec/6_conclusion}
{
    \small
    \bibliographystyle{ieeenat_fullname}
    \bibliography{main}
}

\end{document}

%% file: sec/0_abstract.tex
\begin{abstract}
Audio-visual event localization (AVEL) plays a critical role in multimodal scene understanding.
While existing datasets for AVEL predominantly comprise landscape-oriented long videos with clean and simple audio context, short videos have become the primary format of online video content due to the the proliferation of smartphones. 
Short videos are characterized by portrait-oriented framing and layered audio compositions (\textit{e.g.}, overlapping sound effects, voiceovers, and music), which brings unique challenges unaddressed by conventional methods.
To this end, we introduce \textbf{AVE-PM}, the first AVEL dataset specifically designed for portrait mode short videos, comprising 25,335 clips that span 86 fine-grained categories with frame-level annotations.
Beyond dataset creation, our empirical analysis shows that state-of-the-art AVEL methods suffer an average 18.66\% performance drop during cross-mode evaluation. Further analysis reveals two key challenges of different video formats: 1) spatial bias from portrait-oriented framing introduces distinct domain priors, and 2) noisy audio composition compromise the reliability of audio modality.
To address these issues, we investigate optimal preprocessing recipes and the impact of background music for AVEL on portrait mode videos. Experiments show that these methods can still benefit from tailored preprocessing and specialized model design, thus achieving improved performance.
This work provides both a foundational benchmark and actionable insights for advancing AVEL research in the era of mobile-centric video content.
Dataset and code will be released.
\end{abstract}

%% file: sec/1_intro.tex
\section{Introduction}
\label{sec:intro}

As a pivotal task in multimodal scene understanding, audio-visual event localization (AVEL) has gained significant attention due to its wide-ranging applications. Since the publication of the AVE dataset \cite{tian2018Audiovisual}, considerable progress has been made in this field \cite{xu2020CrossModal, ramaswamy2020What, wang2023Semantic, rao2022Dual, ge2023Learning, xue2023AudioVisual}. 
Recent introductions of diverse datasets including LLP \cite{tian2020Unified}, XD-Violence \cite{wu2020Not} and UnAV-100 \cite{geng2023DenseLocalizing} have further expanded the scope of investigation.

\begin{figure}[t]
  \centering
  \includegraphics[width=\linewidth]{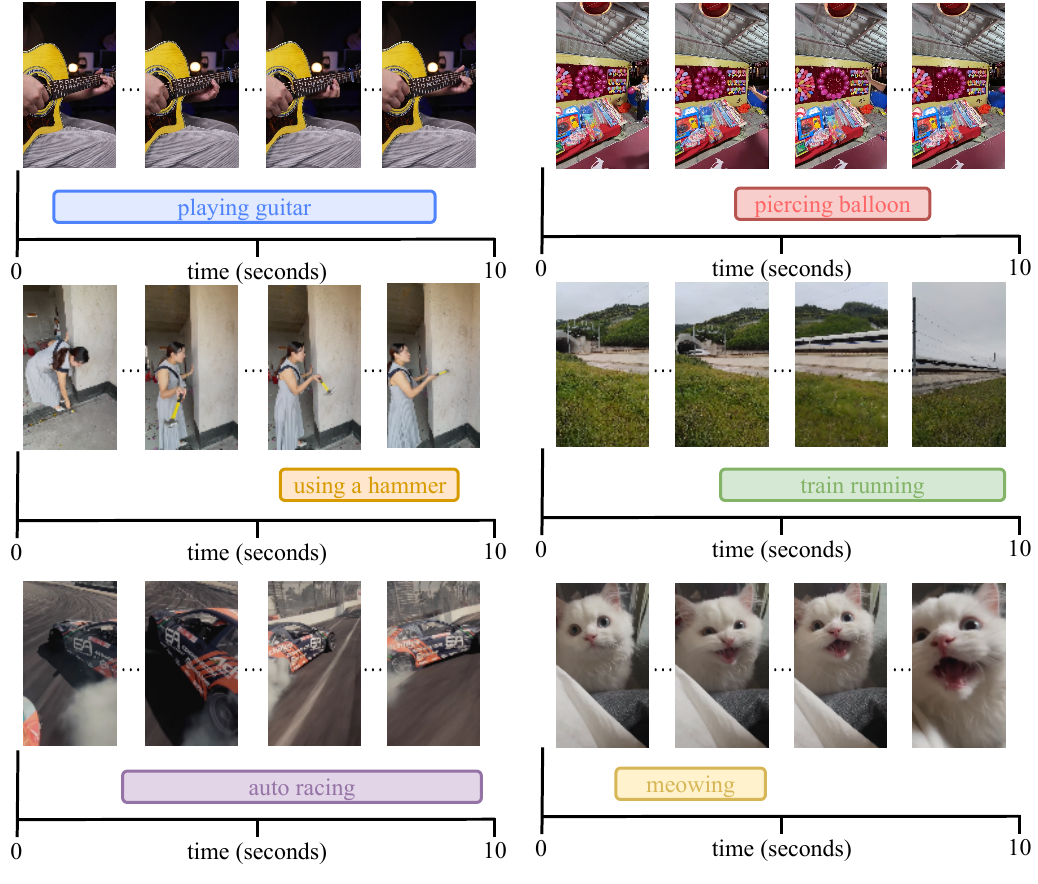}
   \caption{A glance of AVE-PM, the first audio-visual event dataset on short videos with human-annotated temporal boundaries. It consists of 25,335 10-second videos that span over 8 domains and 86 categories. The samples presented here are \textit{playing guitar, piercing balloon, using a hammer, train running, auto racing} and \textit{meowing}.}
   \label{fig:1_glance}
\end{figure}

Contemporary AVEL datasets are predominantly constructed using landscape-oriented long videos sourced from platforms like YouTube \cite{tian2018Audiovisual, tian2020Unified, geng2023DenseLocalizing} and movies \cite{wu2020Not}. However, the proliferation of smartphones and social media has established portrait-oriented short videos as the primary format of online video content \cite{qin2023Content}. 
This transition from landscape mode to portrait mode not only imply a simple change in the aspect ratio, but also brings fundamental changes to user behavior and content characteristics. As demonstrated in \cite{han2024Video}, portrait mode videos exhibit stronger subject focus (typically humans) with reduced background context and increased first-person perspective content. Moreover, users tend to create complex audio compositions featuring layered soundtracks (\textit{e.g.}, overlapping sound effects, voiceovers and music.)
These distinctive characteristics present novel challenges for AVEL systems, as existing methods struggle to generalize to portrait mode videos when trained with landscape mode videos, while the increasingly intricate audio content adds more to the difficulty.

In this paper, we present the \textbf{Audio-visual Event in Portrait Mode (AVE-PM)} dataset, the first portrait mode short video dataset dedicated to AVEL research. The dataset contains 25,335 10-second video clips that span over 86 fine-grained categories with human-annotated event onsets and offsets. Detailed illustration of AVE-PM is presented in \cref{fig:avepm_stats}. All the videos are sourced from Douyin \footnote{Douyin is a popular social media application built for smartphones and primarily features portrait mode short-form videos. \url{https://www.douyin.com/}}, ensuring authentic representation of unconstrained user-generated content comparable to the AVE dataset \cite{tian2018Audiovisual}.

In this study, we extend our investigation beyond dataset creation to address three critical research problems in audio-visual event localization (AVEL) on portrait mode short videos:
\begin{enumerate}
    \item Can existing AVEL models trained on landscape mode datasets generalize to portrait mode videos, and vice versa? For a rigorous comparison, we selected 10 overlapping categories from AVE dataset \cite{tian2018Audiovisual} and AVE-PM, constructing two subsets: Selected-LM and Selected-PM. We conducted cross-mode evaluations with multiple state-of-the-art AVEL models. An average 18.66\% performance drop demonstrates significant degradation in all the selected models, which reveals the domain gap between landscape and portrait mode videos.
    \item What are the fundamental differences between landscape mode and portrait mode videos? From the perspective of AVEL tasks, we identified two key aspects: 1) the influence of spatial bias in the video domain, and 2) the complexity of audio content. We validated these issues by visualizing accuracy heatmaps and measuring the contribution score of both modalities. These findings further emphasize the necessity of studying AVEL in short videos.
    \item Are there effective strategies to mitigate aforementioned problems? To tackle spatial bias, we investigate multiple preprocessing recipes to capture diverse visual information and emphasize the importance of random cropping for better performance. To reveal the impact of complex audio composition, we evaluate selected AVEL methods by excluding training videos with background music and reveal that specialized model designs ensure robust learning even interfered by audio noise, indicating the necessity of further exploration into portrait mode videos.
\end{enumerate}

%% file: sec/2_related_works.tex
\begin{table}[ht]
  \centering
  \small
  \begin{tabular}{@{}lccccc@{}}
    \toprule
    Dataset & Type & Videos & Classes & Length & EB \\
    \midrule
    Audioset \cite{gemmeke2017Audio} & LM & 2.1M & 527 & 10s & \ding{55} \\
    PM-400 \cite{han2024Video} & PM & 76k & 400 & 27s & \ding{55} \\ 
    \midrule
    AVE \cite{tian2018Audiovisual} & LM & 4,143 & 28 & 10s & \ding{51} \\
    LLP \cite{tian2020Unified} \ & LM & 11,849 & 25 & 10s & \ding{51} \\
    XD-Violence \cite{wu2020Not} & LM & 4,754 & 6 & 2.74m & \ding{51} \\
    UnAV-100 \cite{geng2023DenseLocalizing} & LM & 10,790 & 100 & 42.1s & \ding{51} \\
    \midrule
    AVE-PM (Ours) & PM & 25,335 & 86 & 10s & \ding{51} \\
    \bottomrule
  \end{tabular}
  \caption{Comparison with related audio-visual datasets. LM: landscape mode. PM: portrait mode. EB: event boundaries.}
  \label{tab:dataset_comp}
\end{table}

\section{Related work}
\label{sec:formatting}

\subsection{Audio-visual event datasets}

Large-scale audio-visual datasets like Kinetics-Sound \cite{arandjelovic2017Look}, AudioSet \cite{gemmeke2017Audio} and VGGSound \cite{chen2020Vggsound} contribute to advancing audio-visual learning and recognition tasks in machine perception. However, these datasets only contains clip-level annotations with event boundaries. Audio-visual event localization (AVEL) is more intricate because it requires both classification and localization of audio-visual events. AVE dataset \cite{tian2018Audiovisual} is the first AVEL dataset, which is a subset of AudioSet \cite{gemmeke2017Audio} with event temporal boundaries annotated. LLP dataset \cite{tian2020Unified} introduced audio-visual event parsing where video samples contains multiple events. UnAV-100 dataset \cite{geng2023DenseLocalizing} proposed dense localization of multiple audio-visual events in untrimmed videos. Aforementioned datasets are all sourced from landscape-oriented videos, while recent research has focused on developing datasets and methods for audio-visual recognition in diverse video formats, especially short videos in portrait mode. 3MASSIV \cite{gupta20223MASSIV} is a multilingual and multimodal dataset of short social media videos which includes a great proportion of portrait mode videos. However, it focuses on visual concepts rather than specific actions, with only 34 coarse concepts in total. PortraitMode-400 (PM-400) \cite{han2024Video}, the first dataset consisting of portrait mode short videos for action recognition, has addressed challenges unique to this format. Detailed comparison with related audio-visual datasets is shown in \cref{tab:dataset_comp}.

\begin{figure*}
  \centering
  \begin{subfigure}{\linewidth}
    \includegraphics[width=\linewidth]{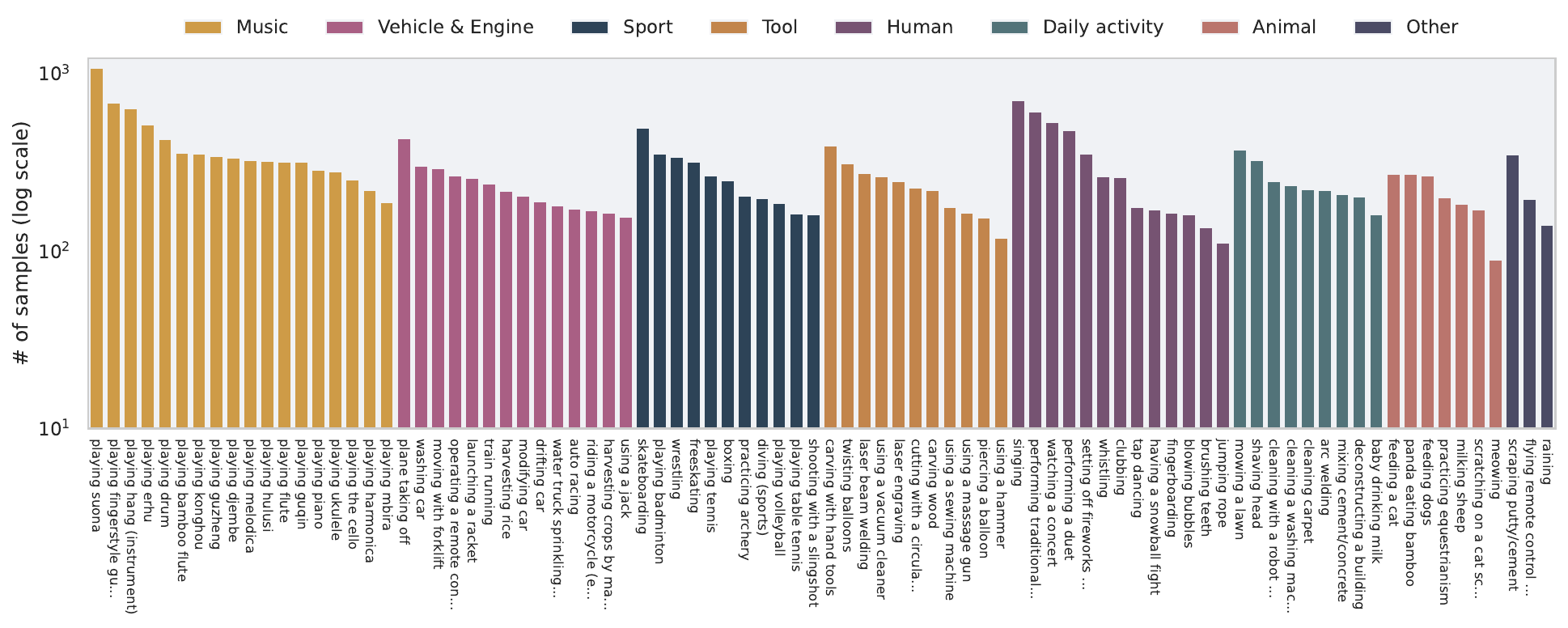}
    \caption{Distribution of events per category}
    \label{fig:avepm_stats-a}
  \end{subfigure}
  \hfill
  \begin{subfigure}{0.5\linewidth}
    \includegraphics[width=\linewidth]{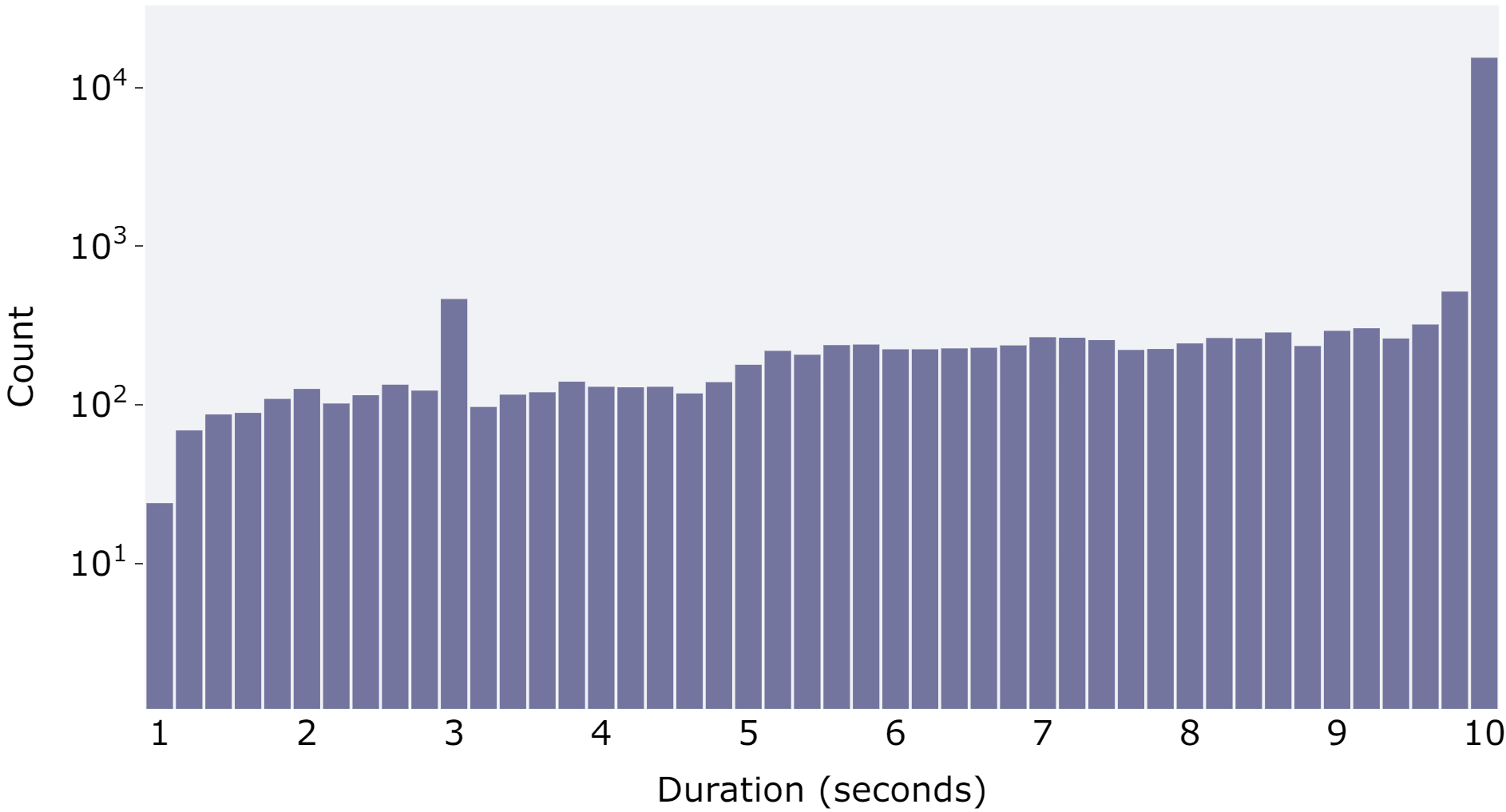}
    \caption{Distribution of event duration}
    \label{fig:avepm_stats-b}
  \end{subfigure}
  \hfill
  \begin{subfigure}{0.3\linewidth}
    \includegraphics[width=\linewidth]{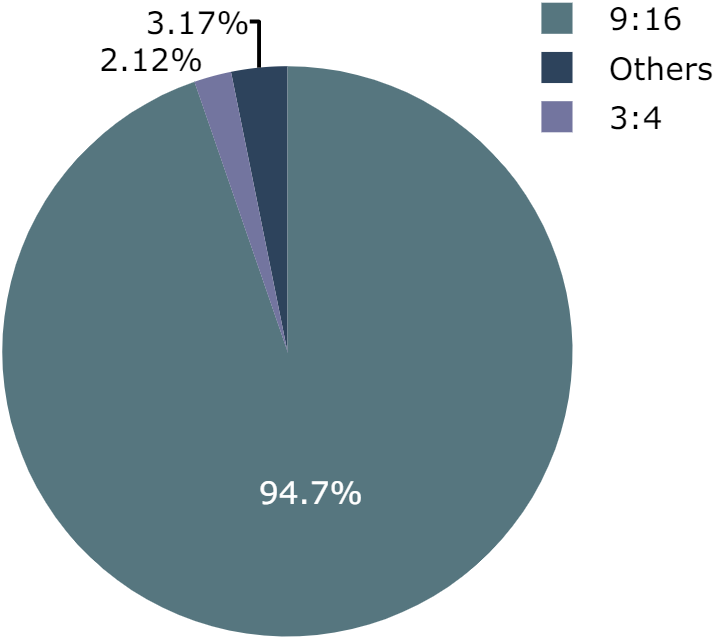}
    \caption{Distribution of aspect ratios (width:height)}
    \label{fig:avepm_stats-c}
  \end{subfigure}
  \caption{Illustrations of statistics on AVE-PM. (a) Distribution of number of events per category. Categories are grouped by domains. Different colors represent different domains. (b) Distribution of event duration. (c) Distribution of aspect ratios in AVE-PM, where 94.7\% videos are in portrait mode with 9:16 format (width:height).}
  \label{fig:avepm_stats}
\end{figure*}

\subsection{Audio-visual event localization}

Recent advances in audio-visual event localization focus on enhancing cross-modal alignment and temporal modeling. Attention mechanisms have been widely adopted, including bidirectional global-local attention \cite{wu2019Dual} and cross-modal co-attention \cite{lin2020Audiovisual, xuan2020CrossModal}. 
To address modality interactions, AVSDN \cite{lin2019Dualmodality} applies a sequence-to-sequence cross-modal architecture, while relation-aware networks \cite{xu2020CrossModal} and semantic modulation frameworks \cite{wang2023Semantic} explicitly model audio-visual correlations. Special architectures like MM-Pyramid \cite{yu2022MMPyramid} and MPN \cite{yu2021MPN} leverage multi-scale features, while \cite{wu2022Spanbased} improve event continuity modeling with span-based approaches. Weakly-supervised methods address label scarcity via contrastive learning strategies \cite{zhou2021Positive, zhou2023Contrastive} and novel loss functions \cite{xue2023AudioVisual}. To mitigate noise interference, \cite{xia2022Crossmodal} adopts background suppression techniques. Optimization methods like OGM-GE \cite{peng2022Balanced} alleviate modality imbalance. Recent innovations also explore efficient adaptation of pre-trained vision transformers \cite{lin2023Vision} and latent summarization for temporal inconsistency \cite{he2021Multimodal, feng2024CSSNet}. However, since all the available datasets are mainly constructed with clips from landscape-oriented long videos, the ability to generalize on portrait mode videos have not yet been discussed.

%% file: sec/3_ave_pm.tex
\section{The AVE-PM dataset}
\label{sec:3_dataset}
In this section, we describe the process of build AVE-PM dataset and provide statistical analysis of its data distribution. First, we begin with the introduction of its taxonomy. Next, we describe the data collection and annotation process. Finally, we provide statistical analysis on event durations and specific categories.

\subsection{Taxonomy}
\label{sec:taxonomy}

Following the practice of AVE dataset \cite{tian2018Audiovisual}, the most commonly used dataset in AVEL, we select specific categories from PortraitMode-400 \cite{han2024Video}, the first dataset dedicated to portrait mode video recognition. The hierarchical tree structure taxonomy of PortraitMode-400 is then inherited in AVE-PM. Although most of the videos in PortraitMode-400 contain an audio track, not all the categories precisely match the common definition of audio-visual events (e.g., \textit{Makeup} and \textit{performing acupuncture}). Therefore, we built the ontology graph of PortraitMode-400 and compared it with the ontology of AudioSet \cite{gemmeke2017Audio} to obtain 200 candidate categories in PortraitMode-400 that possibly contain audio-visual events. Then, we randomly sampled 20 videos from each category and provided them to expert annotators as a test run, where we filtered out 100 candidate categories for further annotation. Finally, we regrouped these categories into 8 high-level domains that covers most of the occasions in daily life, spanning from human activities to natural sounds as shown in \cref{fig:avepm_stats-a}.

\subsection{Dataset construction}
\label{sec:dataset_construction}

\subsubsection{Data collection}
\label{sec:data_collection}
According to the video ids provided in \cite{han2024Video}, we collected raw videos from Douyin platform, a popular social media application built for smartphones and primarily features portrait mode short videos. We performed an audio quality analysis and observed that a large proportion of videos contain background musics. While previous datasets such as AVE \cite{tian2018Audiovisual} and UnAV-100 \cite{geng2023DenseLocalizing} have excluded such videos to ensure clean audio tracks, we argue that this approach may not fully align with real-world scenarios, as background music is prevalent in short videos. Removing these videos alters the data distribution, thus limiting the potential for further applications. Therefore, we choose to provide a \texttt{haveBGM} flag for each annotated video so that the quality of the dataset is guaranteed while utilizing these noisy videos remains an option.

\subsubsection{Data annotation}
We developed a custom video annotation tool for clearer visualization and annotated raw videos by crowdsourcing. Presented with the category of target audio-visual event, annotators are asked to mark the onset and offset of the event on the waveform graph of provided video, as well as confirm the presence of background music within the region. To facilitate accurate temporal boundary identification, which can be challenging based solely on visual cues, annotators are provided with both the waveform and spectrogram of the audio track. To ensure annotation quality, approximately 20\% of videos have at least two annotations from two different annotator. A third annotation is required if two annotations differs two much (\textit{i.e.}, a discrepancy of 0.5 seconds or more in either the onset or offset) from each other. 

\subsubsection{Post processing}
Since the durations of raw videos vary from 8 seconds to 1 minute, we cut the raw videos into multiple 10-second clips, following the practice of AVE dataset \cite{tian2018Audiovisual}. We then discard the clips in which the event lasts less than 1 second. We also filtered out the categories where valid clips are less than 100, resulting in discarding 14 categories out of 100 annotated categories. Through post processing, we managed to guarantee that each category contains at least 114 clips.

\begin{figure}[ht]
  \centering
  \includegraphics[width=\linewidth]{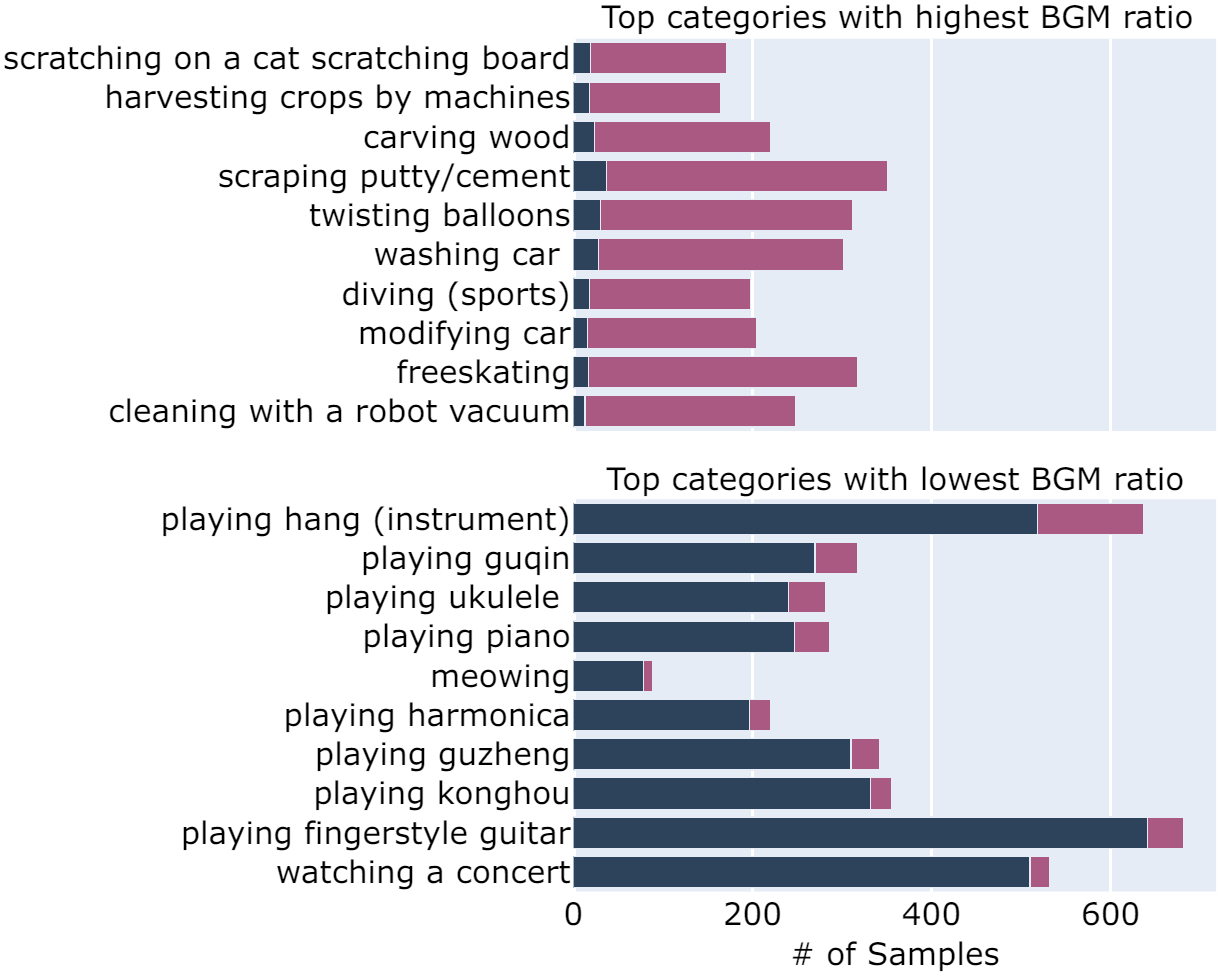}
   \caption{Distribution of categories with the highest and lowest BGM ratios. The top subplot shows the top 10 categories with the highest BGM ratio, while the bottom subplot displays the top 10 categories with the lowest BGM ratio. The bars represent the count of samples with and without BGM for each category.}
   \label{fig:bgm_ratio}
\end{figure}

\subsection{Statistical analysis}

In detail, AVE-PM dataset contains 24,450 10-second video clips, where each clips contains one single audio-visual event with its temporal onset and offset annotated. We split the dataset into training, validation and testing sets with a ratio of 6:2:2. The samples from each category are distributed into each subset according to this ratio, thereby guaranteeing consistency in the data distribution across the subsets. The illustrations of statistics are presented in \cref{fig:avepm_stats}.

In \cref{fig:bgm_ratio}, we present the proportion of samples containing background music for each category. The tendency of users to add background music varies significantly depending on the content of the videos. For instance, in the categories of \textit{meowing} and \textit{playing fingerstyle guitar}, the proportion of samples with background music is less than 10\% of the total samples in each category. In contrast, in the categories of \textit{freeskating} and \textit{skateboarding}, this proportion exceeds 80\%. Although the target events remain audible, the presence of background music still poses a challenge for accurate event localization.

%% file: sec/4_experiments.tex
\begin{table*}[ht]
    \centering
    \begin{tabular}{@{}lccccccc@{}}
        \toprule
        Method & \makecell{Visual \\ Encoder} & \makecell{Audio \\ Encoder} & \makecell{Visual Pretrain \\ Dataset} & \makecell{Audio Pretrain \\ Dataset} & \makecell{Total Params \\ (M)} & \makecell{Acc. on\\ AVE} & \makecell{Acc. on\\ AVE-PM} \\
        \midrule
        AVELN \cite{tian2018Audiovisual}   & VGG-19    & VGGish & ImageNet & AudioSet & 136.0 & 74.0$^{\dagger}$ & 71.19 \\
        CPSP \cite{zhou2023Contrastive}  & VGG-19        & VGGish & ImageNet & AudioSet & 217.4 & 77.8$^{\dagger}$ & 75.79 \\
        CMBS \cite{xia2022Crossmodal}  & VGG-19    & VGGish & ImageNet & AudioSet & 315.2 & 79.7$^{\dagger}$ & 77.99 \\
        LAVISH \cite{lin2023Vision} & \multicolumn{2}{c}{Swin-V2-L (Shared)} & ImageNet & \ding{55} & 238.8 & \textbf{81.1}$^{\dagger}$ & \textbf{79.37} \\
        \bottomrule
    \end{tabular}
    \caption{Localization accuracy (\%) of selected methods on AVE and AVE-PM dataset. $\dagger$ indicates that the results are from corresponding papers where the encoders are pretrained. Results on AVE-PM are our runs.}
    \label{tab:methods_comparison}
\end{table*}

\section{Cross-mode audio-visual event localization}
\label{sec:cross_mode_avel}

This section investigates the distinct characteristics of portrait mode videos within the context of audio-visual event localization. Driven by user behavior and device constraints, landscape mode and portrait mode videos exhibit inherent differences in spatial priors and audio composition. We hypothesize that models trained on one orientation may struggle to generalize to the other due to these intrinsic biases. To validate this, we conduct a comprehensive cross-mode evaluation, followed by analysis on spatial priors and audio complexity. 

\subsection{Experiment setup}
To ensure a rigorous comparison between landscape mode and portrait mode audio-visual event localization, we select 10 overlapping categories from the 28 classes of the AVE dataset to construct a subset. We utilize all samples from the corresponding categories of the AVE dataset to build the AVE subset landscape mode (S-LM), which comprises 1,536 samples, accounting for 37\% of the total 4,143 samples in the AVE dataset. Subsequently, we select an equal number of samples per category from the AVE-PM dataset to construct the AVE-PM subset portrait mode (S-PM). By ensuring identical taxonomy and equal data distribution per category across both subsets, we establish a fair testing condition to validate the differences in audio-visual event localization between landscape and portrait videos, where the primary distinction between the subsets lies in the data content itself.

We selected four distinct methods for comparison to encompass a diverse range of network architectures. \textbf{AVELN} \cite{tian2018Audiovisual} is a dual multimodal residual network designed for the joint modeling of auditory and visual clues. \textbf{CPSP} \cite{zhou2023Contrastive} employs a contrastive positive sample propagation method to enhance feature representation learning. \textbf{CMBS} \cite{xia2022Crossmodal} is a cross-modal background suppression network aimed at reducing noise and improving localization performance. These models all adopt separate visual and audio encoders, utilizing pre-trained VGG and VGGish networks to extract video and audio features. In a different direction, \textbf{LAVISH} \cite{lin2023Vision} explores the use of a pretrained Swin transformer \cite{liu2022Swin}, introducing a latent audio-visual hybrid adapter that achieves competitive performance with fewer tunable parameters. We report detailed information on the selected models and their audio-visual event localization performance on both AVE and AVE-PM under standard fully supervised training recipe in \cref{tab:methods_comparison} for reference.


\begin{table}[ht]
    \centering
    \begin{tabular}{lccccc}
        \toprule
        Method & Train & Test & Acc. & \makecell{Acc.\\drop} & \makecell{Avg.\\Acc.} \\
        \midrule
        \multirow{4}{*}{AVELN \cite{tian2018Audiovisual}} & \multirow{2}{*}{LM} & LM & \textbf{70.42} & \multirow{2}{*}{\textcolor{red}{-20.55}} & \multirow{2}{*}{60.14} \\ 
                              &                      & PM & 49.87 & & \\ \cmidrule{2-6}
                              & \multirow{2}{*}{PM} & LM & 59.65 & \multirow{2}{*}{\textcolor{red}{-11.64}} & \multirow{2}{*}{65.47} \\ 
                              &                      & PM & \textbf{71.29} & & \\ \midrule
        \multirow{4}{*}{CPSP \cite{zhou2023Contrastive}} & \multirow{2}{*}{LM} & LM & \textbf{73.83} & \multirow{2}{*}{\textcolor{red}{-21.32}} & \multirow{2}{*}{63.17} \\ 
                              &                      & PM & 52.51 & & \\ \cmidrule{2-6}
                              & \multirow{2}{*}{PM} & LM & 65.53 & \multirow{2}{*}{\textcolor{red}{-7.88}} & \multirow{2}{*}{69.47} \\ 
                              &                      & PM & \textbf{73.41} & & \\ \midrule
        \multirow{4}{*}{CMBS \cite{xia2022Crossmodal}} & \multirow{2}{*}{LM} & LM & \textbf{73.36} & \multirow{2}{*}{\textcolor{red}{-11.01}} & \multirow{2}{*}{67.87} \\ 
                              &                      & PM & 62.36 & & \\ \cmidrule{2-6}
                              & \multirow{2}{*}{PM} & LM & 50.34 & \multirow{2}{*}{\textcolor{red}{\textbf{-25.41}}} & \multirow{2}{*}{63.04} \\ 
                              &                      & PM & \textbf{75.74} & & \\ \midrule
        \multirow{4}{*}{LAVISH \cite{lin2023Vision}} & \multirow{2}{*}{LM} & LM & \textbf{85.85} & \multirow{2}{*}{\textcolor{red}{\textbf{-21.77}}} & \multirow{2}{*}{\textbf{74.97}} \\ 
                                &                      & PM & 64.08 & & \\ \cmidrule{2-6}
                                & \multirow{2}{*}{PM} & LM & 74.41 & \multirow{2}{*}{\textcolor{red}{-12.64}} & \multirow{2}{*}{\textbf{80.72}} \\ 
                                &                      & PM & \textbf{87.04} & & \\ \bottomrule
    \end{tabular}
    \caption{Cross-mode evaluation accuracy (\%) of selected methods on S-LM and S-PM.}
    \label{tab:cross_mode_eval}
\end{table}

\subsection{Cross-mode evaluation}
\label{sec:cross_eval}
To demonstrate the domain differences between landscape mode (LM) and portrait mode (PM) videos in the context of audio-visual event localization, we conducted a cross-mode evaluation on the S-LM and S-PM subsets. We trained the selected models on different subsets and evaluated their performance on the test sets of both subsets, as shown in \cref{tab:cross_mode_eval}.

From the experimental results, the first observation we can make is that all models exhibit their best performance when trained and tested on the same subset. This indicates that audio-visual event localization in portrait mode videos is not a trivial problem that can be simply addressed by training existing models on current AVE datasets and directly applying them to portrait mode videos. This suggests that training with portrait mode videos is necessary for existing methods to be applied to diverse scenarios like localizing audio-visual events in short videos.

Another observation is that all models show varying degrees of performance degradation in cross-mode evaluation. Among the models trained on the S-LM subset and tested on the S-PM subset, LAVISH experiences the largest accuracy drop, with a decrease on accuracy of 21.77\%. Conversely, among the models trained on the S-PM subset and tested on the S-LM subset, CMBS shows the largest accuracy drop, with a decrease of on accuracy 25.41\%. This implies that there are significant domain prior differences between portrait mode and landscape mode videos, and existing methods are not effectively designed to generalize between these two modes. Therefore, further research is needed to address the unique characteristics of portrait mode data.

\subsection{Analysis on spatial priors}
We aim to further investigate the underlying reasons for the observed performance differences between landscape and portrait videos, with the hypothesis that different spatial priors exist between landscape and portrait videos. To validate this hypothesis, we employ a sliding window approach to investigate the impact of different regions within the video frames on the overall accuracy for both formats.

For this experiment, we select LAVISH \cite{lin2023Vision}, the top-performing method from our cross-evaluation in \cref{sec:cross_eval}, for this study. Training videos from S-LM and S-PM are randomly resized with a shorter-side length between 440 and 512, then center-cropped to 192$\times$192. During evaluation, a sliding window generates 192$\times$192 crops from test videos in both subsets from various locations within the frames. These crops are passed to the model to obtain region-specific evaluation results. For portrait videos (S-PM), the stride is set to 1/9 of the width and 1/16 of the height, reflecting the 9:16 aspect ratio, which is prevalent in this subset as indicated in \cref{fig:avepm_stats-c}. For landscape videos (S-LM), the stride is adjusted to match the 16:9 aspect ratio.

\begin{figure}[ht]
  \centering
  \includegraphics[width=\linewidth]{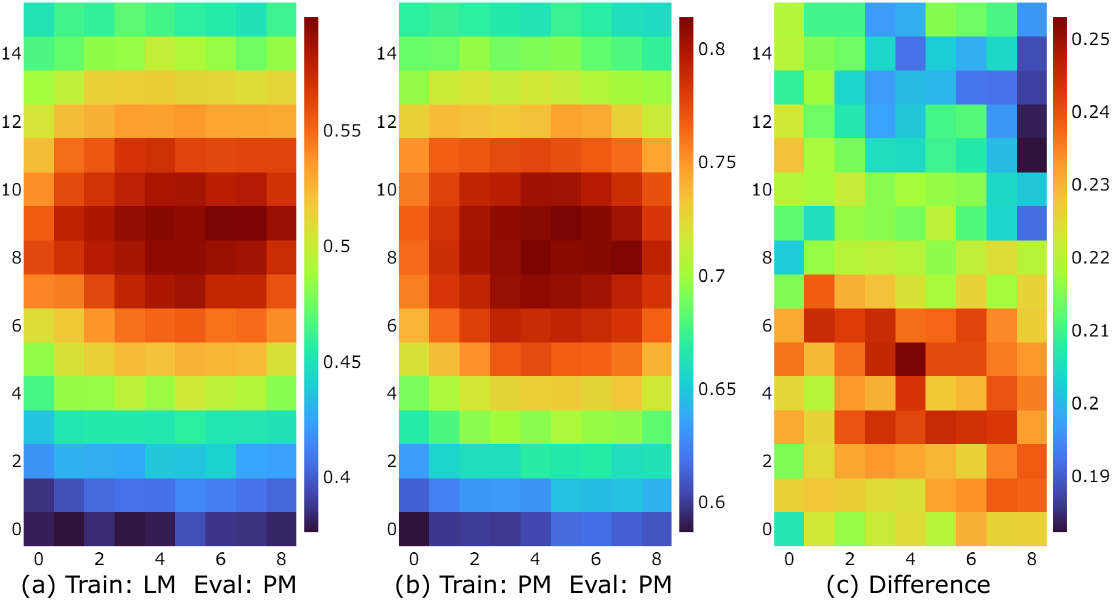}
    \vspace{-0.5em}
   \caption{The accuracy heatmaps of evaluating LAVISH at different spatial locations on the S-PM subset. (a) Accuracy heatmap of LAVISH model trained on S-LM. (b) Accuracy heatmap of LAVISH model trained on S-PM. (c) The difference map represents the subtraction of the accuracy of the model trained on S-LM from the model trained on S-PM, \textit{i.e.}, (b) - (a).}
   \label{fig:acc_map_pm}
\end{figure}

\begin{figure}[ht]
  \centering
  \includegraphics[width=\linewidth]{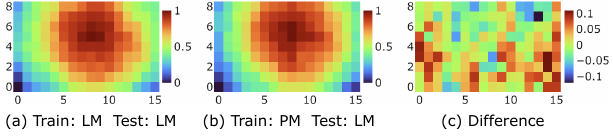}
   \caption{The accuracy heatmaps of evaluating LAVISH at different spatial locations on the S-LM subset. (a) Accuracy heatmap of LAVISH model trained on S-LM. (b) Accuracy heatmap of LAVISH model trained on S-PM. (c) The difference map represents the subtraction of the accuracy of the model trained on S-PM from the model trained on S-LM, \textit{i.e.}, (a) - (b).}
   \label{fig:acc_map_lm}
\end{figure}

With the accuracies from different regions, we compose accuracy heatmaps of all the four different train-eval scenarios as well as the difference heatmap. When conducting evaluation on S-PM subset, the sliding strategy results in two 9$\times$16 heatmaps from the model trained on S-LM and S-PM, as shown in \cref{fig:acc_map_pm}. For evaluation on S-LM subset, the corresponding heatmaps are 16$\times$9, as shown in \cref{fig:acc_map_lm}. Each position on the heatmap represents the average accuracy on that region, while the difference heatmap shows the corresponding accuracy differences at each region. The difference at each region represents the accuracy difference of the same model when trained on S-LM and S-PM.

From the heatmaps of each evaluation scenario, it is evident that the highest accuracy consistently occurs at the frame center, regardless of video orientation, indicating that most audio-visual event information is centralized. However, as presented in Table \ref{tab:cross_mode_eval}, the model trained on S-PM achieves a higher overall accuracy compared to the model trained on S-LM (87.04\% \textit{vs.} 64.08\%). As shown in Figure \ref{fig:acc_map_pm} (c), this phenomenon can be attributed to the fact that the informative area in portrait videos is more concentrated in the lower half of the frame, and the model trained on S-PM effectively captures this unique data distribution pattern. The bottom region of portrait videos contains visual priors associated with the events, which is the fundamental reason for the suboptimal performance of the model trained on S-LM.

We also present the accuracy heatmaps of S-LM evaluations in \cref{fig:acc_map_lm} (a) and (b), where the model trained on S-LM achieves a higher overall accuracy than the model trained on S-PM (85.85\% \textit{vs.} 74.41\%). From the difference between these two heatmaps in \cref{fig:acc_map_lm} (c), it is observed that the informative areas on the sides of landscape videos are the primary cause of this performance disparity. In real-world scenarios, the sides of landscape videos typically encompass richer environmental information, leading to differences in visual priors between landscape and portrait videos.

\subsection{Analysis on audio composition}
\label{sec:audio_analysis}
Due to user behavior and device constraints, short video creators often add a significant amount of artificial sound effects, voiceover, and background music before uploading a video, which can sometimes completely obscure the event information in the audio track. In such cases, utilizing the audio data not only fails to capture event information but may also interfere with the video modality.

To this end, we introduce modality contribution score proposed in \cite{mai2024Multimodal} to evaluate the model's reliance on information from a particular modality during classification. For the audio modality $a$ and video modality $v$, the modality contribution scores are defined as:

\begin{equation}
        \text{mcs}_i = \frac{1}{l_i + \gamma_i}
\label{eq:mcs}
\end{equation}
where $i \in \{a, v\}$ and $l_i$ is the predictive loss from corresponding modality. The hyperparameter $\gamma_i$ serves as a scaling factor that ensures the denominator remains non-zero and stabilizes the scores when the predictive loss is small or approaches zero.

As described in \cref{sec:data_collection}, each video clip has a boolean annotation \texttt{haveBGM}, describing whether the clip contains background music. Note that \texttt{haveBGM} being true indicates that background music is present in the event region, but the target event is still audible.

To investigate the impact of audio complex, we train LAVISH \cite{lin2023Vision} on S-PM subset with both video and audio information from all training videos. Then, we obtain $\text{mcs}_a$ and $\text{mcs}_v$ from the videos in the test split of these two subsets. 
We conducted a point-biserial correlation analysis between the provided \texttt{haveBGM} annotations and calculated modality contribution scores to examine whether the audio contribution score changes when the audio contains background music, and whether there is a correlation between these two scenarios.

The point-biserial correlation analysis reveals significant negative correlations between \texttt{haveBGM} status and $\text{mcs}_a$ across multiple event categories, with correlation coefficients ranging from -0.68 (\emph{train running}) to -0.33 (\emph{playing harmonica}). This indicates that the presence of background music substantially reduces the model's reliance on audio information, particularly for events with characteristic sound patterns. In contrast, $\text{mcs}_v$ show weak correlations, suggesting minimal compensatory reliance on visual information when audio quality degrades.

Notably, categories requiring fine-grained audio discrimination (\emph{auto racing}, -0.46) exhibit stronger negative correlations than those with distinctive visual patterns (\emph{flying remote control drone}, 0.03). The exception of \emph{playing flute} (0.01) may stem from its unique spectral characteristics that survive musical interference. The lack of significant positive correlations in $\text{mcs}_v$ implies current multimodal architectures fail to effectively redistribute attention between modalities when one becomes unreliable, highlighting the need for adaptive fusion mechanisms in audio-visual event understanding.

\begin{figure}[ht]
  \centering
  \includegraphics[width=\linewidth]{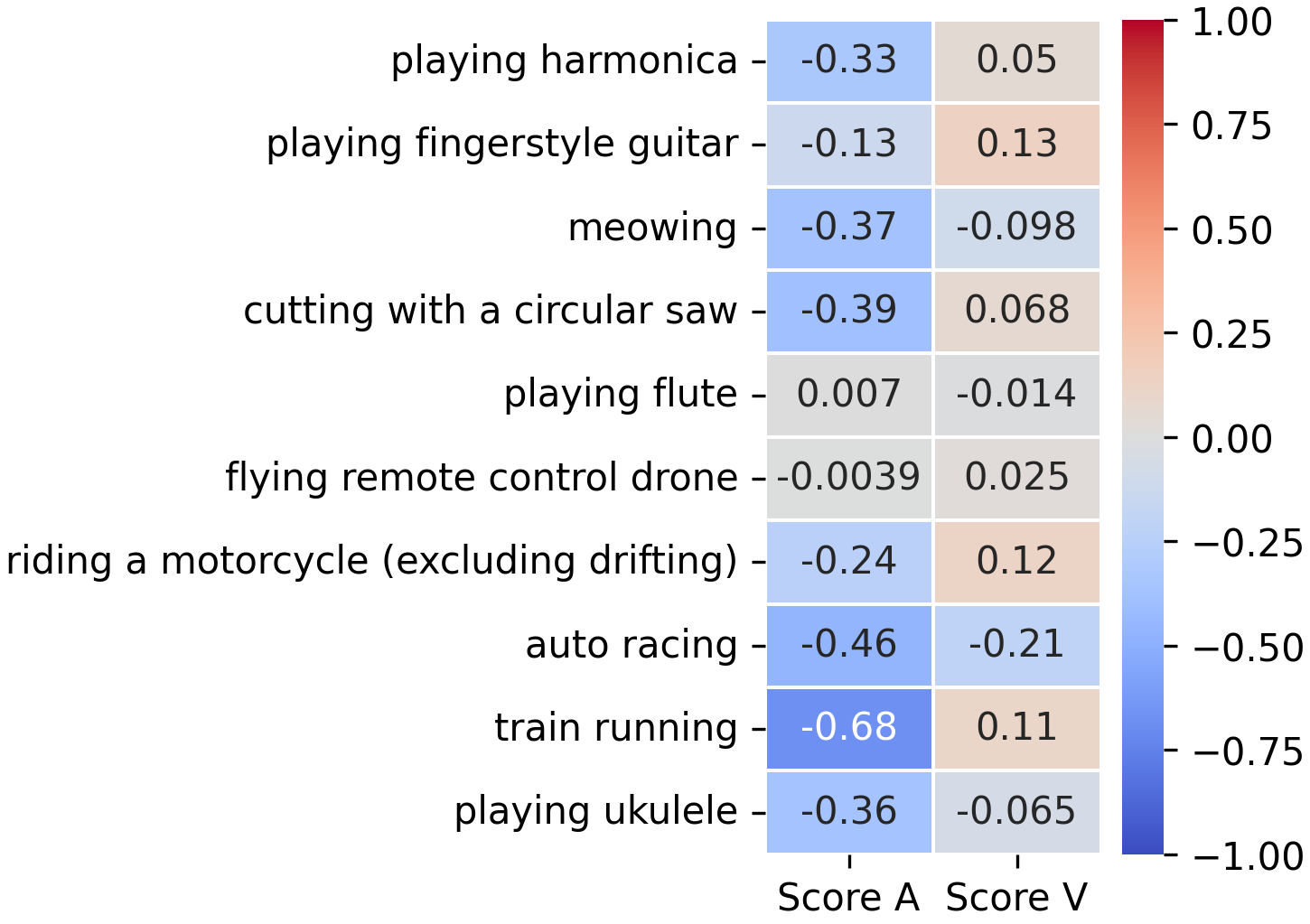}
  \vspace{-0.5em}
   \caption{Distribution of modality contribution scores on each category in S-PM.}
   \label{fig:score_distribution_by_category}
\end{figure}

%% file: sec/5_recipes.tex
\section{Importance of data preprocessing}

In the preceding section, we demonstrated through cross-mode evaluation that the distinct data priors of landscape mode and portrait mode videos pose novel challenges for existing audio-visual event localization methods.
To address these challenges, in this section, we aim to mitigate the biases introduced by their unique format and identify the best preprocessing recipes for portrait mode audio-visual event localization.

\subsection{Resizing and cropping}

Due to the unconstrained nature of the videos in AVE and AVE-PM dataset, resizing and cropping are necessary preprocessing steps for a consistent input size and aspect ratio. Most of the audio-visual event localization methods utilize VGG network \cite{simonyan2015Very} pretrained on ImageNet \cite{deng2009ImageNet} for visual feature extraction, like AVELN \cite{tian2018Audiovisual}, CPSP \cite{zhou2023Contrastive}, CMBS \cite{xia2022Crossmodal} and other methods \cite{wang2024ContextAware, he2021Multimodal, ramaswamy2020What, xuan2021Discriminative, wu2019Dual, yu2022MMPyramid, xue2023AudioVisual, feng2024CSSNet}. Therefore, the center cropping size of these methods are set to 224$\times$224 to match the input size of VGG. Recently proposed vision transformer \cite{dosovitskiy2021Image, liu2022Swin} based methods like LAVISH \cite{lin2023Vision} and other methods \cite{chalk2024TIM, lin2024Siamese} utilize patch-embedded visual frames in the shape of 192$\times$192 as direct input instead of using VGG features. Before conducting center cropping on input frames, all aforementioned methods either adopt shorter-side resizing to keep the original aspect ratio or simply resize the visual frame to a square shape of 224$\times$224 or 192$\times$192.

\begin{table}[ht]
    \centering
    \begin{tabular}{lcccc}
        \toprule
               & \multirow{2}{*}{Orig.} & \multicolumn{2}{c}{Shorter.} & \multirow{2}{*}{Incep.} \\ \cmidrule(lr){3-4}
               & &  center & random & \\ 
        \midrule
        AVELN \cite{tian2018Audiovisual}  & 71.29 & 74.02 & 72.89 & \textbf{74.98} \\
        CPSP \cite{zhou2023Contrastive}  & 73.41 & 74.60 & \textbf{77.59} & 77.04 \\
        CMBS \cite{xia2022Crossmodal}    & 75.74 & 75.72 & 76.91 & \textbf{77.62} \\
        LAVISH \cite{lin2023Vision}      & \textbf{87.04} & 86.30 & 86.56 & 86.01 \\
        \bottomrule
    \end{tabular}
    \caption{Comparison of accuracy (\%) for different preprocessing strategies applied to portrait mode videos. ``Orig.'' denotes the original preprocessing pipeline from the selected methods. ``Short.'' represents shorter-side resizing, followed by either center cropping or random cropping. ``Incep.'' refers to Inception-style resizing, which incorporates random sampling, cropping, and resizing.}
    \vspace{-10pt}
    \label{tab:comp_resizing}
\end{table}

In this subsection, we investigate the effectiveness of popular video preprocessing strategies for audio-visual event localization, \textit{i.e.}, shorter-side resizing method \cite{simonyan2015Very} and the Inception-style method \cite{feichtenhofer2019slowfast, fan2021multiscale, li2023uniformer}. Shorter-side resizing involves resizing the shorter side of the frame to a fixed length or a random value within a range \cite{wang2021tdn} while scaling the longer side proportionally. Inception-style method, on the other hand, involves random sampling, resizing and cropping, which generates a more diverse group of inputs. 

As shown in \cref{tab:comp_resizing}, the experimental results reveal distinct preprocessing preferences across methods. VGG-based methods achieve their best performance with Inception-style resizing (74.98\% for AVELN and 77.62\% for CMBS) or shorter-side resizing with random cropping (77.59\% for CPSP), indicating that random operations enhance robustness to aspect ratio distortions. In contrast, Vision Transformer-based LAVISH performs best with its original preprocessing (87.04\%) where it directly resize the frames to $192\times192$ without keeping its original aspect ratio. Shorter-side resizing slightly degrades LAVISH’s performance (-0.48\% to -0.74\%), indicating its sensitivity to aspect ratio changes. These findings suggest that portrait mode videos require specialized preprocessing strategies. Inception-style methods enhance traditional CNNs by introducing diversity, while further investigations are required for deciding the best preprocessing recipe for ViT-based methods.

\begin{table}[ht]
    \centering
    \begin{tabular}{@{}lcc@{}}
    \toprule
    & BGM included & BGM excluded \\ \midrule
    AVELN \cite{tian2018Audiovisual}   & 71.29 & \textbf{72.60} \\ 
    CPSP \cite{zhou2023Contrastive}   & 73.41 & \textbf{75.76} \\
    CMBS \cite{xia2022Crossmodal}   & \textbf{75.74} & 72.70 \\
    LAVISH \cite{lin2023Vision} & \textbf{87.04} & 85.40 \\ \bottomrule
    \end{tabular}
    \caption{Comparison of model performance on audio-visual event localization with and without excluding training videos containing background music (BGM). The table highlights the accuracy (\%) for each model under both conditions, demonstrating the impact of BGM on performance.}
    \label{tab:haveBGM}
    \vspace{-5pt}
\end{table}

\subsection{Excluding videos with background music}

As discussed in \cref{sec:audio_analysis}, short videos in AVE-PM contain a significant amount of artificial sound effects, voiceover, and background musics. To investigate the impact of background music (BGM) on audio-visual event localization, we conducted experiments to determine whether excluding videos with BGM during training improves model performance by reducing audio interference. We utilized the \texttt{haveBGM} annotation in S-PM subset, where each video clip is annotated with a boolean flag indicating BGM presence, as mentioned in \cref{sec:data_collection}. The models are trained and evaluated on S-PM subset under two conditions: (1) using all training data, and (2) excluding clips with BGM in trainning data. 

The experimental results reveal distinct model behaviors in handling background music (BGM) during audio-visual event localization. AVELN and CPSP show performance improvements (1.31\% and 2.35\% respectively) when excluding BGM, validating the hypothesis that non-event-related audio can cause modal interference. In contrast, CMBS, with its dedicated cross-modal background suppression network \cite{xia2022Crossmodal}, achieves a significant 3.04\% performance boost with BGM, validating the effectiveness of its background suppression mechanism in noise reduction. LAVISH, on the other hand, achieve superior performance (1.64\% higher with BGM) with robust feature extraction capabilities of the pretrained swin transformer \cite{liu2022Swin}. 
The results suggest that the videos with background music still contain useful information for audio-visual event localization, but specialized model designs are required to effectively utilize this information.

%% file: sec/6_conclusion.tex
\section{Conclusion}
In this paper, we introduce the Audio-visual Event in Portrait Mode (AVE-PM) dataset, the first dataset dedicated to audio-visual event localization in portrait mode short videos. Through comprehensive experiments, we demonstrated that existing AVEL models struggle to generalize across video modes, revealing a significant domain gap. We also identify the key differences between landscape mode and portrait mode videos, such as spatial bias and audio complexity, highlighting the need for specialized approaches. We make initial attempts to investigate optimal preprocessing techniques like random cropping, and present potential approaches  to mitigate audio noise. We hope AVE-PM provides a foundation for future research, encouraging further research on portrait mode videos.